\pdfoutput=1
\documentclass[conference]{IEEEtran}
\usepackage[utf8]{inputenc}
\usepackage{algorithm}
\usepackage{algpseudocode}
\usepackage{amsmath,amssymb,amsfonts, amsthm}
\usepackage{balance}
\usepackage{caption}
\usepackage{cite}
\usepackage{color}
\usepackage{comment}
\usepackage{calc}
\usepackage{float}
\usepackage{multirow}
\usepackage{nicefrac}
\usepackage{enumerate}
\usepackage{textcomp}
\usepackage{verbatim}
\usepackage{xcolor}
\usepackage{lipsum}  
\usepackage{units}
\usepackage{subcaption}
\usepackage{orcidlink}
\usepackage{pgfplots}
\usepackage{tikz}
\DeclareUnicodeCharacter{2212}{−}
\pgfplotsset{width=10cm,compat=1.9}
\usepgfplotslibrary{groupplots,dateplot}
\usetikzlibrary{patterns,shapes.arrows}
\pgfplotsset{compat=newest}

\def\BibTeX{{\rm B\kern-.05em{\sc i\kern-.025em b}\kern-.08em
    T\kern-.1667em\lower.7ex\hbox{E}\kern-.125emX}}

\title{\textit{IoTScent}: Enhancing Forensic Capabilities in Internet of Things Gateways}

\author{\IEEEauthorblockN{Antonio Boiano\, \orcidlink{0000-0002-5552-3680}}
\IEEEauthorblockA{\textit{DEIB, Politecnico di Milano} \\
Milan, Italy \\
antonio.boiano@polimi.it}
\and
\IEEEauthorblockN{Alessandro Enrico Cesare Redondi }
\IEEEauthorblockA{\textit{DEIB, Politecnico di Milano} \\
Milan, Italy \\
alessandroenrico.redondi@polimi.it
}
\and
\IEEEauthorblockN{Matteo Cesana }
\IEEEauthorblockA{\textit{DEIB, Politecnico di Milano} \\
Milan, Italy \\
matteo.cesana@polimi.it
}
}

\date{30 June 2023}

\begin{document}

\maketitle

\begin{abstract}
The widespread deployment of Consumer Internet of Things devices in proximity to human activities makes them digital observers of our daily actions.
This has led to a new field of digital forensics, known as IoT Forensics, where digital traces generated by IoT devices can serve as key evidence for forensic investigations. Thus, there is a need to develop tools that can efficiently acquire and store network traces from IoT ecosystems.
This paper presents IoTScent, an open-source IoT forensic tool that enables IoT gateways and Home Automation platforms to perform IoT traffic capture and analysis. Unlike other works focusing on IP-based protocols, IoTScent is specifically designed to operate over IEEE 802.15.4-based traffic, which is the basis for many IoT-specific protocols such as Zigbee, 6LoWPAN and Thread. 
IoTScent offers live traffic capture and feature extraction capabilities, providing a framework for forensic data collection that simplifies the task of setting up a data collection pipeline, automating the data collection process, and providing ready-made features that can be used for forensic evidence extraction. 
This work provides a comprehensive description of the IoTScent tool, including a practical use case that demonstrates the use of the tool to perform device identification from Zigbee traffic. The study presented here significantly contributes to the ongoing research in IoT Forensics by addressing the challenges faced in the field and publicly releasing the IoTScent tool.
\end{abstract}


\section{Introduction}
\label{sec:introduction}
Considering the Internet of Things (IoT) ecosystem, Consumer IoT (CIoT) devices play a significant role, as they are becoming increasingly prevalent in our society.
According to \cite{cisco2020cisco}, it is projected that CIoT devices will have the largest share, accounting for approximately 50\% of the entire IoT device market.
This is also demonstrated by the investment made by the Connectivity Standard Alliance, composed of major CIoT hardware and software manufacturers in the smart home field, e.g., Apple, Amazon, and Google, into developing new protocols such as Thread and Matter.
The widespread deployment of CIoT devices in homes, workplaces, and buildings where people spend most of their time makes them natural witnesses of daily human activities. Indeed, CIoT devices are designed to interact and support human actions (e.g., controlling lights/doors and electronic equipment); it follows that the behaviour of CIoT devices is strictly connected with human daily lives and can be monitored and analyzed to perform tasks such as human activity recognition. Recently, IoT forensics has emerged as a new research area focusing on identifying and extracting digital information from IoT devices. In particular, the network traffic exchanged by such devices may be conveniently captured, stored and analyzed to provide evidence that can be used in a court of justice \cite{IoT_for_rev}. 
Moreover, monitoring and analysing the network traffic produced by CIoT devices is key for protecting user's security and privacy: the generally poor security design of CIoT devices, due to manufacturers' priorities on cost reduction and time to market, makes them prone to a vast range of cyberattacks \cite{8977812}.
In general, enabling a-posteriori network traffic analysis in CIoT scenario is a challenging task due to the fragmentation inherent in IoT communication technologies (edge/cloud), the transient and ephemeral nature of the traffic generated by IoT devices, and the limited memory size of these devices.

All these observations call for tools able to identify, collect and preserve IoT network traffic to support IoT forensic analyses. Most works dealing with IoT traffic collection and analyses focus on traffic traces at the IP layer or leverage Wi-Fi traffic \cite{FABIO_OWRT}\cite{Inform_exposure_iot_mandalari}. Although many currently available CIoT devices are Wi-Fi-based, an increasing number of devices use other communication technologies based on low-power, short-range standards such as IEEE 802.15.1 (Bluetooth) or IEEE 802.15.4 (Zigbee/Thread). It is crucial to develop tools and methodologies to capture and analyse such traffic efficiently. Moreover, networks based on such standards generally rely on the presence of specific hub devices which aggregate traffic and bridge it to the Internet using WiFi/Ethernet connectivity. Examples include popular proprietary solutions such as Philips Hue Bridge and open-source platforms such as Home Assistant\footnote[1]{https://github.com/home-assistant}. These intermediate points may shadow important details in the network traffic exchanged locally, calling for solutions which can monitor traffic closer to the generating devices. 

This paper presents \textit{IoTScent} (IoTS), an IoT forensic tool which enables forensic analysis on IEEE 802.15.4 networks controlled by the open-source Home Assistant platform. IoTS enables Home Assistant to perform several tasks related to IoT forensics, such as traffic capture and feature extraction, providing the basis for network monitoring and analysis tasks. The architecture and implementation of IoTS are presented, highlighting its features through a practical use case relative to IoT device identification. 

The subsequent sections of this paper are structured as follows: Section~\ref{c:rel_work} offers a presentation of the primarily related studies in IoT forensics. Section~\ref{c:tool} presents an accurate description of the IoTScent tool, describing its architecture, capabilities, and performance. Section~\ref{c:use_case} presents a  practical use-case where Device Identification is performed over features extracted by IoTScent. The study concludes in Section~\ref{c:conc}, which provides final reflections and highlights directions for future research.

\section{Related Work}
\label{c:rel_work}
Over the past few years, there has been a surge of interest in CIoT, leading to the emergence of a subfield within Digital Forensics, namely IoT Forensics. Numerous studies have been conducted in this field with the aim of defining IoT Forensics enabler frameworks.
Cicirelli et al., \cite{Cicirelli2016} proposed a framework for developing a human activity recognition system in a smart home environment. Zawoad et al., \cite{Zawoad2015} instead propose a theoretical framework for IoT Device Forensics, IoT Network Forensics, and IoT Cloud Forensics. 
However, there has been limited research into developing and defining practical frameworks for IoT forensic data collection similar to the one presented in this paper. It is noteworthy that Meffert et al., \cite{FSAIoT} proposed a centralized Forensic State Acquisition Controller framework, which simultaneously performs state collection and control of IoT devices. Unlike the solution presented in this article, the framework does not directly monitor encrypted traffic for state collection but rather records logs received directly from IoT devices and end-user interaction.
On the other hand, Palmese et al., \cite{FABIO_OWRT} proposed and developed Feature-Sniffer, a user-friendly tool that computes network traffic features on OpenWrt-based Wi-Fi access points. The work shares similarities in technique and collected features compared with the tool proposed in this paper. However, while Wi-Fi-based traffic can provide valuable insights into network usage and behaviour, Feature-Sniffer does not include monitoring and analysing traffic from 802.15.4 networks which may limit its ability to detect certain types of devices or behaviours.
\section{Tool Overview} \label{c:tool}

\begin{figure}
\centerline{\includegraphics[width=0.43\textwidth]{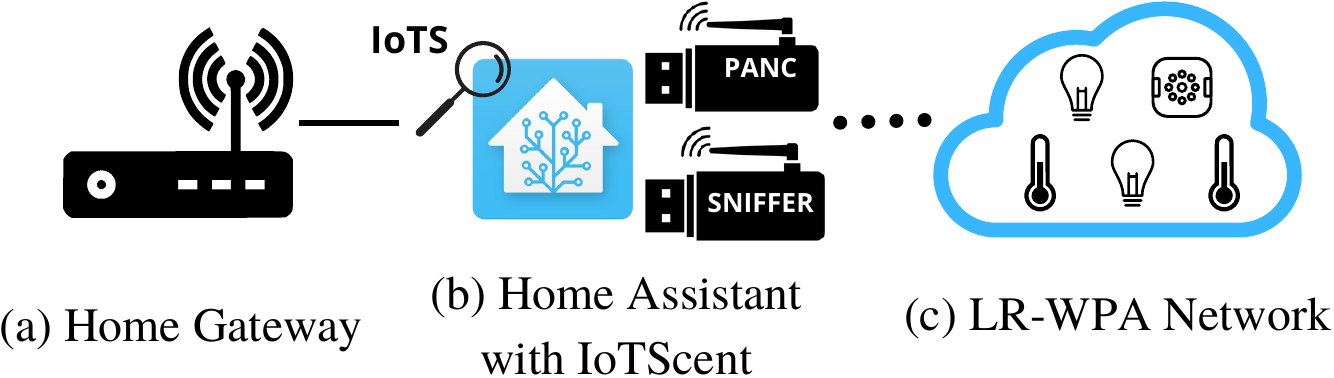}}
\caption{IoTScent Application Setup}
\label{fig:arc}
\end{figure}

\begin{figure*}
\centering
\includegraphics[width=0.66\textwidth]{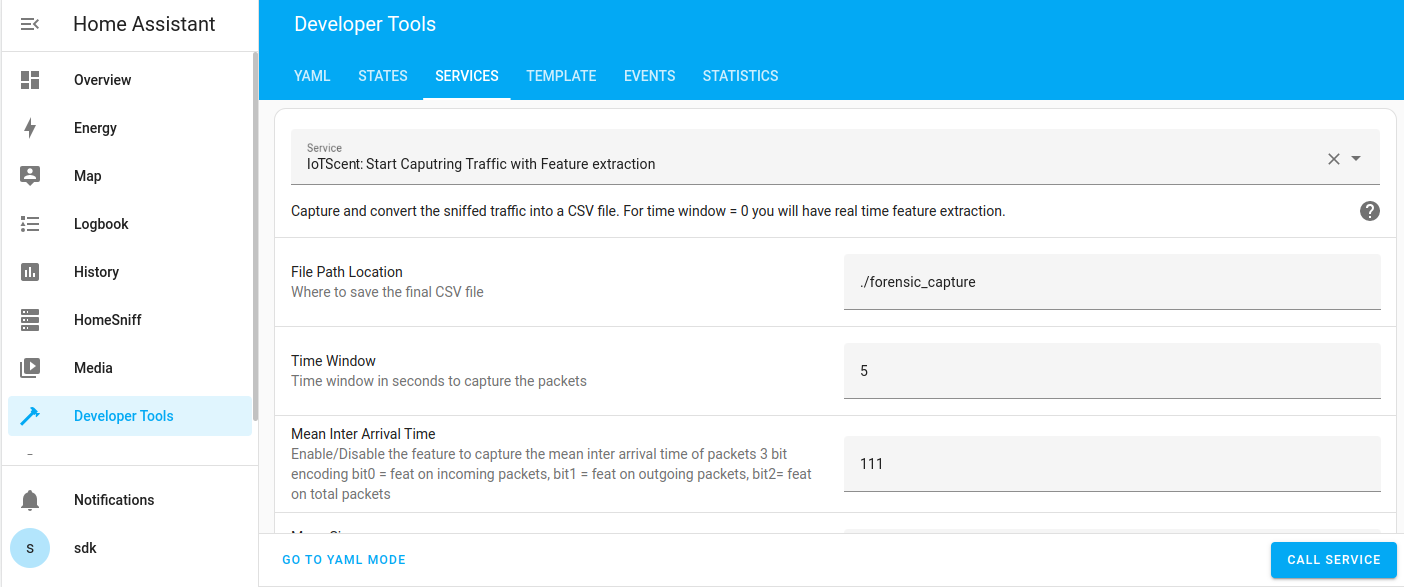}
\caption{IoTScent feature extraction as a service call, Home Assistant User Interface}
\label{fig:ui}
\end{figure*}

\textit{IoTScent}\footnote[2]{https://github.com/antonio-boiano/IoTScent/} is an open-source tool, publicly available on GitHub, designed for forensic analysis in networks operated by the IEEE 802.15.4 standard. The tool is implemented as a custom integration to the Home Assistant OS, which can be operated on low-cost general-purpose hardware such as the Raspberry PI. IoTScent eases the tedious task of setting up and automating the general IoT forensic data collection pipeline and provides ready-to-use functionalities for capturing network traffic and extracting its characteristics in a few steps. Traffic capture is performed through the availability of a dedicated radio interface compatible with the Killerbee Software suite,
which was integrated into the Home Assistant framework. Compatible solutions include, e.g., Crossbow Telosb, Texas Instruments CC2530/1, and Silicon Labs EFR32 family SoC.
A lightweight dissector is implemented to convert raw data streams into formatted MAC and Zigbee Network packet headers.

In a nutshell, IoTScent provides the following functionalities:
\begin{itemize}
    \item \textit{Packet capture:} IoTS allows for TCPdump-like traffic capture over the 802.15.4 physical layer, saving raw packet data in a PCAP file format. Traditional packet filters can be used to filter incoming packets (e.g., source/destination addresses, packet type, PAN address, etc.), enabling processing only a subset of sniffed packets.
    \item \textit{Time-windowed feature extraction:} Most IoT forensic analysis tasks are based on statistical features extracted from network traffic. IoTS allows easily extracting network features in a time-windowed fashion from 802.15.4 traffic. In detail, IoTS organizes the received traffic in time windows of user-defined duration and computes a set of traffic features for each time window and for each device, identified by its source address present in the 802.15.4 MAC header. The output of the feature extraction processes is saved in a CSV file and stored inside the device for easy management and post-processing of the retrieved information. To provide the network administrator with informative data that can be used for IoT forensics analysis, the IoTS feature extraction capabilities have been refined through a systematic review of the most commonly used network features for device identification, classification, and human activity recognition in low-rate wireless personal area (LR-WPA) networks \cite{Acar2020},\cite{iot_mosaic},\cite{Babun2020}. IoTS employs three primary feature components: Packet Size, Payload Length, and Inter-arrival Time (IAT) to conduct statistical analysis over user-defined time windows. 
    This analysis involves extracting the Mean and Standard Deviation values from the aforementioned features. To improve the computational performance, the mean and the standard deviation are computed online using Welford's formulation \cite{Welford}. 
    Furthermore, IoTS allows users to differentiate incoming and outgoing packets and perform separate or combined statistics for each direction.
    Finally, IoTS also permits extracting features from an existing PCAP file containing 802.15.4 raw packets, allowing offline traffic analysis.
\end{itemize}

\subsection{IoTScent Overview}
\subsubsection{Hardware setup}
IoTS can be seamlessly deployed on both IoT gateways or other devices running the Home Automation platform. Such devices generally have no energy or performance constraints and are well-suited for executing network traffic analysis and feature extraction operations.
A standard implementation setup for IoTS is shown in Figure~\ref{fig:arc}. In this configuration, a Raspberry Pi is utilized as an IoT gateway, running the Home Assistant platform, and an IEEE 802.15.4 antenna (SONOFF USB Dongle based on EFR32MG21 SoC) is used as the main Personal Area Network (PAN) Coordinator (PANC) hardware. IoTS is installed as an integration on the Home Assistant platform and requires little configuration (PANC channel) and a compatible IEEE 802.15.4 antenna. The tool was tested with a USB dongle based on the CC2531 System on Chip (SoC), used to perform packet sniffing in promiscuous mode.

\subsubsection{Software setup}
The IoTS controlling logic comprises two primary components: the first is responsible for accessing hardware resources, performing traffic capture and distributing the captured raw packets among asynchronous queues, whereas the second component is dedicated to performing analysis and feature extraction from the packet queues. By separating these tasks, IoTS can avoid resource conflict while efficiently handling the data produced by IoT networks.
The two components are detailed in the following:
\begin{enumerate}[i)]
    \item {\textit{Packet capture:}} Upon IoTS startup, a single task is initiated to manage the 802.15.4 SoC hardware component. This design choice enables the necessary hardware configurations and capability checks to be performed and prevents concurrent task operations, which could result in deadlocks. Multiple asynchronous queues with limited buffers can be defined, allowing multiple capture tasks to run in parallel and limiting IoTS's impact on computational resources. This prevents packet loss in packet burst scenarios and enables each feature extraction task to be handled independently. Limiting the queue size is a prevention mechanism in case the hardware on which IoTS is installed cannot handle the packet rate produced by the observed network. To limit the impacts of IoTS on normal IoT gateway execution, different mechanics are implemented. IoTS can be configured to stop the feature extraction process or limit the packet rate (with packet loss) when the queue reaches its maximum capacity. Additionally, prior to the packet distribution task, this layer is also responsible for packet filtering based on the rules defined by each acquisition process. The syntax used by IoTS for specifying packet filters is based on the one used by Wireshark for Zigbee and 802.15.4 packets.
    \item {\textit{ Traffic Feature extraction:}} The second IoTS component is dedicated to processing the raw packet data received from the first component. Upon launching a new acquisition task, an asynchronous queue is created, which is then passed to the first component to populate it with raw packet data. In addition, it extracts only the user-selected relevant features from the raw packet data to optimise CPU and RAM consumption. These features are then used to calculate the user-requested features and are saved in CSV for the live feature extraction process. Alternatively, the extracted features are saved to a second queue, which will be processed depending on the time windows to extract the requested features per device.
\end{enumerate}

\subsubsection{User Interface}
The IoTS capabilities have been exposed as the so-called Home Assistant Services, which allow the network administrator to interact with IoTS in a wide range of scenarios easily; indeed, service calls can be directly called from Home Assistant's user interface (UI) under the Developer Tools menu, or triggered by Home Assistant Automation script, which allows scheduling or triggering the feature extraction processes. With this approach, network administrators can easily interact with IoTS functionality.
System administrators opting for the IoTScent services will encounter five distinct service implementations designed to handle the IoTScent execution tasks: (i) start and configure the Features extraction task, (ii) TCPdump-like traffic capture, (iii) termination of an acquisition task, (iv) retrieval of its execution status, and (v) removal of an acquisition task from memory. Figure~\ref{fig:ui} illustrates a UI section related to the IoTS feature extraction service call. On this page, users can specify the output file's destination path, time window duration, and the set of features they want IoTS to compute.


\begin{table}
\centering
\caption{CIoT devices utilized as a testbed in the study}
\label{tab0}
\begin{tabular}{|c|c|c|c|}
\hline
\textbf{ID} & \textbf{Device Model} & \textbf{Brand} & \textbf{Actions} \\ \hline
(a) & Hue Motion Sensor & Philips & \begin{tabular}[c]{@{}c@{}}Motion\\  Light Intensity\end{tabular} \\ \hline
(b) & TS011F & Tuya & \begin{tabular}[c]{@{}c@{}}On \& Off\\  Power Consumption\end{tabular} \\ \hline
(c) & Plug Z3 & Ledvance & \begin{tabular}[c]{@{}c@{}}On \& Off\\  Power Consumption\end{tabular} \\ \hline
(d) & Hue White Lamp & Philips & \begin{tabular}[c]{@{}c@{}}On \& Off\\ Luminosity\end{tabular} \\ \hline
(e) & Door Window Sensor & Aqara & Open \& Closed \\ \hline
(f) & ZBSA-Motion Sensor & Woolley & Motion \\ \hline
(g) & TS0043 Switch & Tuya & \begin{tabular}[c]{@{}c@{}}Short press\\ Long press\\ Double press\end{tabular} \\ \hline
\end{tabular}%
\end{table}

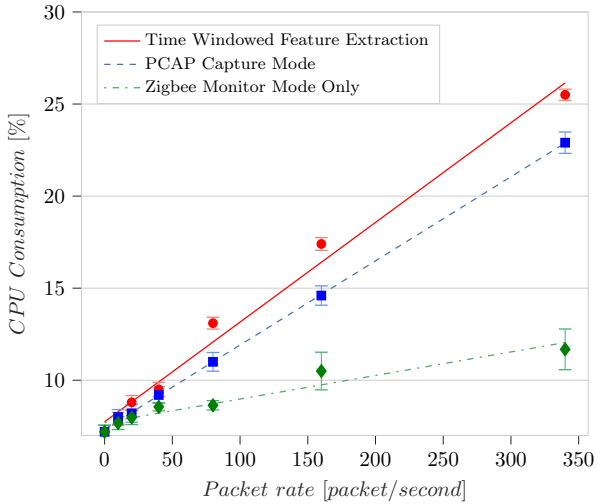
\begin{figure}
\centering
\begin{tikzpicture}[scale=0.80]

\definecolor{cornflowerblue}{RGB}{100,149,237}
\definecolor{darkslategray38}{RGB}{38,38,38}
\definecolor{green}{RGB}{0,128,0}
\definecolor{indianred1967882}{RGB}{196,78,82}
\definecolor{lightgray204}{RGB}{204,204,204}
\definecolor{mediumseagreen}{RGB}{60,179,113}
\definecolor{mediumseagreen85168104}{RGB}{85,168,104}
\definecolor{salmon}{RGB}{250,128,114}
\definecolor{steelblue76114176}{RGB}{76,114,176}

\begin{axis}[
axis line style={lightgray204},
legend cell align={left},
legend style={
  fill opacity=0.8,
  draw opacity=1,
  text opacity=1,
  at={(0.03,0.97)},
  anchor=north west,
  draw=lightgray204,
  font = \footnotesize
},
ytick pos=left,
xtick pos=bottom,
tick align=outside,
x grid style={lightgray204},
xlabel=\textcolor{darkslategray38}{\(\displaystyle Packet\ rate\ [packet/second]\)},
xmajorticks=true,
xmin=-17, xmax=357,
xtick style={color=darkslategray38},
y grid style={lightgray204},
ylabel=\textcolor{darkslategray38}{\(\displaystyle CPU\ Consumption\ [\%]\)},
ymajorgrids,
ymajorticks=true,
ymin=7, ymax=30,
ytick style={color=darkslategray38}
]
\path [draw=salmon]
(axis cs:0,6.84541253648033)
--(axis cs:0,7.55458746351967);

\path [draw=salmon]
(axis cs:10,7.543)
--(axis cs:10,8.257);

\path [draw=salmon]
(axis cs:20,8.43)
--(axis cs:20,9.17);

\path [draw=salmon]
(axis cs:40,9.113)
--(axis cs:40,9.887);

\path [draw=salmon]
(axis cs:80,12.78)
--(axis cs:80,13.42);

\path [draw=salmon]
(axis cs:160,17.053)
--(axis cs:160,17.747);

\path [draw=salmon]
(axis cs:340,25.1921754001971)
--(axis cs:340,25.8078245998029);

\path [draw=cornflowerblue]
(axis cs:0,6.84541253648033)
--(axis cs:0,7.55458746351967);

\path [draw=cornflowerblue]
(axis cs:10,7.591)
--(axis cs:10,8.409);

\path [draw=cornflowerblue]
(axis cs:20,7.719)
--(axis cs:20,8.681);

\path [draw=cornflowerblue]
(axis cs:40,8.757)
--(axis cs:40,9.643);

\path [draw=cornflowerblue]
(axis cs:80,10.492)
--(axis cs:80,11.508);

\path [draw=cornflowerblue]
(axis cs:160,14.075)
--(axis cs:160,15.125);

\path [draw=cornflowerblue]
(axis cs:340,22.322)
--(axis cs:340,23.478);

\path [draw=mediumseagreen]
(axis cs:0,6.84541253648033)
--(axis cs:0,7.55458746351967);

\path [draw=mediumseagreen]
(axis cs:10,7.32279678060588)
--(axis cs:10,8.02720321939412);

\path [draw=mediumseagreen]
(axis cs:20,7.59279678060588)
--(axis cs:20,8.35720321939412);

\path [draw=mediumseagreen]
(axis cs:40,8.33615455555208)
--(axis cs:40,8.77495655555903);

\path [draw=mediumseagreen]
(axis cs:80,8.3874763052236)
--(axis cs:80,8.8925236947764);

\path [draw=mediumseagreen]
(axis cs:160,9.47843438147359)
--(axis cs:160,11.5215656185264);

\path [draw=mediumseagreen]
(axis cs:340,10.5725356358596)
--(axis cs:340,12.783019919696);

\addplot [semithick, salmon, mark=-, mark size=3, mark options={solid}, only marks, forget plot]
table {%
0 6.84541253648033
10 7.543
20 8.43
40 9.113
80 12.78
160 17.053
340 25.1921754001971
};
\addplot [semithick, salmon, mark=-, mark size=3, mark options={solid}, only marks, forget plot]
table {%
0 7.55458746351967
10 8.257
20 9.17
40 9.887
80 13.42
160 17.747
340 25.8078245998029
};
\addplot [semithick, cornflowerblue, mark=-, mark size=3, mark options={solid}, only marks, forget plot]
table {%
0 6.84541253648033
10 7.591
20 7.719
40 8.757
80 10.492
160 14.075
340 22.322
};
\addplot [semithick, cornflowerblue, mark=-, mark size=3, mark options={solid}, only marks, forget plot]
table {%
0 7.55458746351967
10 8.409
20 8.681
40 9.643
80 11.508
160 15.125
340 23.478
};
\addplot [semithick, mediumseagreen, mark=-, mark size=3, mark options={solid}, only marks, forget plot]
table {%
0 6.84541253648033
10 7.32279678060588
20 7.59279678060588
40 8.33615455555208
80 8.3874763052236
160 9.47843438147359
340 10.5725356358596
};
\addplot [semithick, mediumseagreen, mark=-, mark size=3, mark options={solid}, only marks, forget plot]
table {%
0 7.55458746351967
10 8.02720321939412
20 8.35720321939412
40 8.77495655555903
80 8.8925236947764
160 11.5215656185264
340 12.783019919696
};
\addplot [semithick, red]
table {%
0 7.74565078349856
10 8.28688839142949
20 8.82812599936041
40 9.91060121522226
80 12.075551646946
160 16.4054525103933
340 26.14772945315
};
\addlegendentry{$\mathrm{Time\ Windowed\ Feature\ Extraction}$}
\addplot [semithick, steelblue76114176, dashed]
table {%
0 7.3418132395267
10 7.79884873680844
20 8.25588423409018
40 9.16995522865366
80 10.9980972177806
160 14.6543811960345
340 22.8810201471058
};
\addlegendentry{$\mathrm{PCAP\ Capture\ Mode}$}
\addplot [semithick, mediumseagreen85168104, dash pattern=on 1pt off 3pt on 3pt off 3pt]
table {%
0 7.70188226201898
10 7.82973083892975
20 7.95757941584053
40 8.21327656966209
80 8.72467087730519
160 9.74745949259141
340 12.0487338769854
};
\addlegendentry{$\mathrm{Zigbee\ Monitor\ Mode\ Only}$}
\addplot [semithick, red, mark=*, mark size=2, mark options={solid}, only marks, forget plot]
table {%
0 7.2
10 7.9
20 8.8
40 9.5
80 13.1
160 17.4
340 25.5
};
\addplot [semithick, blue, mark=square*, mark size=2, mark options={solid}, only marks, forget plot]
table {%
0 7.2
10 8
20 8.2
40 9.2
80 11
160 14.6
340 22.9
};
\addplot [semithick, green, mark=diamond*, mark size=3, mark options={solid}, only marks, forget plot]
table {%
0 7.2
10 7.675
20 7.975
40 8.55555555555556
80 8.64
160 10.5
340 11.6777777777778
};
\end{axis}

\end{tikzpicture}
\caption{CPU usage of HomeAssistant running \textit{IoTScent} under different packet rates and execution behaviors}
\label{fig:cpufull}
\end{figure}

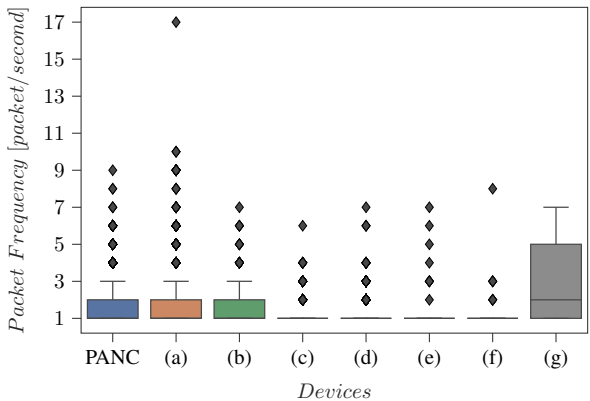
\begin{figure}
\centering
\begin{tikzpicture}[scale=0.8]

\definecolor{darkkhaki193178127}{RGB}{193,178,127}
\definecolor{darkslategray38}{RGB}{38,38,38}
\definecolor{darkslategray73}{RGB}{73,73,73}
\definecolor{gray140}{RGB}{140,140,140}
\definecolor{gray140120102}{RGB}{140,120,102}
\definecolor{indianred1819295}{RGB}{181,92,95}
\definecolor{lightgray204}{RGB}{204,204,204}
\definecolor{lightslategray133122170}{RGB}{133,122,170}
\definecolor{mediumseagreen95157109}{RGB}{95,157,109}
\definecolor{peru20313699}{RGB}{203,136,99}
\definecolor{plum208148190}{RGB}{208,148,190}
\definecolor{steelblue88116163}{RGB}{88,116,163}

\begin{axis}[
ytick pos=left,
xtick pos=bottom,
height=7 cm,
ytick={1,3,5,7,9,11,13,15,17,19},
ymajorticks=true,
xmajorticks=true,
axis line style={darkslategray38},
tick align=outside,
x grid style={lightgray204},
xlabel=\textcolor{darkslategray38}{\(\displaystyle Devices\)},
xmajorticks=true,
xmin=-0.5, xmax=7.5,
xtick style={color=darkslategray38},
xtick={0,1,2,3,4,5,6,7},
xticklabels={PANC,(a),(b),(c),(d),(e),(f),(g)},
y grid style={lightgray204},
ylabel=\textcolor{darkslategray38}{\(\displaystyle Packet\ Frequency\ 
  [packet/second]\)},
ymajorticks=true,
ymin=0.2, ymax=17.8,
ytick style={color=darkslategray38}
]
\path [draw=darkslategray73, fill=steelblue88116163, semithick]
(axis cs:-0.4,1)
--(axis cs:0.4,1)
--(axis cs:0.4,2)
--(axis cs:-0.4,2)
--(axis cs:-0.4,1)
--cycle;
\path [draw=darkslategray73, fill=peru20313699, semithick]
(axis cs:0.6,1)
--(axis cs:1.4,1)
--(axis cs:1.4,2)
--(axis cs:0.6,2)
--(axis cs:0.6,1)
--cycle;
\path [draw=darkslategray73, fill=mediumseagreen95157109, semithick]
(axis cs:1.6,1)
--(axis cs:2.4,1)
--(axis cs:2.4,2)
--(axis cs:1.6,2)
--(axis cs:1.6,1)
--cycle;
\path [draw=darkslategray73, fill=indianred1819295, semithick]
(axis cs:2.6,1)
--(axis cs:3.4,1)
--(axis cs:3.4,1)
--(axis cs:2.6,1)
--(axis cs:2.6,1)
--cycle;
\path [draw=darkslategray73, fill=lightslategray133122170, semithick]
(axis cs:3.6,1)
--(axis cs:4.4,1)
--(axis cs:4.4,1)
--(axis cs:3.6,1)
--(axis cs:3.6,1)
--cycle;
\path [draw=darkslategray73, fill=gray140120102, semithick]
(axis cs:4.6,1)
--(axis cs:5.4,1)
--(axis cs:5.4,1)
--(axis cs:4.6,1)
--(axis cs:4.6,1)
--cycle;
\path [draw=darkslategray73, fill=plum208148190, semithick]
(axis cs:5.6,1)
--(axis cs:6.4,1)
--(axis cs:6.4,1)
--(axis cs:5.6,1)
--(axis cs:5.6,1)
--cycle;
\path [draw=darkslategray73, fill=gray140, semithick]
(axis cs:6.6,1)
--(axis cs:7.4,1)
--(axis cs:7.4,5)
--(axis cs:6.6,5)
--(axis cs:6.6,1)
--cycle;
\addplot [semithick, darkslategray73]
table {%
0 1
0 1
};
\addplot [semithick, darkslategray73]
table {%
0 2
0 3
};
\addplot [semithick, darkslategray73]
table {%
-0.2 1
0.2 1
};
\addplot [semithick, darkslategray73]
table {%
-0.2 3
0.2 3
};
\addplot [black, mark=diamond*, mark size=2.5, mark options={solid,fill=darkslategray73}, only marks]
table {%
0 4
0 4
0 6
0 6
0 6
0 6
0 4
0 4
0 4
0 5
0 5
0 4
0 4
0 4
0 4
0 4
0 4
0 5
0 4
0 6
0 4
0 4
0 4
0 6
0 4
0 4
0 4
0 6
0 5
0 4
0 4
0 4
0 5
0 4
0 4
0 4
0 4
0 4
0 4
0 5
0 4
0 4
0 5
0 4
0 6
0 4
0 4
0 4
0 5
0 4
0 4
0 4
0 5
0 5
0 4
0 4
0 4
0 4
0 4
0 4
0 4
0 6
0 4
0 4
0 4
0 4
0 4
0 4
0 4
0 4
0 4
0 4
0 7
0 6
0 4
0 5
0 4
0 4
0 4
0 5
0 4
0 4
0 4
0 4
0 5
0 6
0 4
0 5
0 5
0 6
0 6
0 4
0 4
0 4
0 5
0 6
0 4
0 4
0 5
0 4
0 4
0 4
0 4
0 4
0 4
0 6
0 4
0 4
0 4
0 6
0 4
0 4
0 6
0 4
0 4
0 4
0 4
0 4
0 4
0 6
0 4
0 4
0 4
0 9
0 5
0 4
0 4
0 4
0 4
0 4
0 4
0 4
0 4
0 6
0 5
0 5
0 4
0 5
0 5
0 4
0 4
0 4
0 8
0 4
0 4
0 4
0 4
0 6
0 4
0 5
0 4
0 4
0 4
0 4
0 4
0 4
0 4
0 5
0 4
0 6
0 4
0 4
0 4
0 4
0 4
0 4
0 4
0 4
0 8
0 4
0 4
0 4
0 4
0 6
0 4
0 4
0 4
0 4
0 5
0 4
0 4
0 7
0 4
0 4
0 4
0 4
0 4
0 4
0 4
0 4
0 4
0 4
0 7
0 5
0 4
0 4
0 4
0 4
0 5
0 6
0 4
0 4
0 4
0 5
0 4
0 4
0 4
0 4
};
\addplot [semithick, darkslategray73]
table {%
1 1
1 1
};
\addplot [semithick, darkslategray73]
table {%
1 2
1 3
};
\addplot [semithick, darkslategray73]
table {%
0.8 1
1.2 1
};
\addplot [semithick, darkslategray73]
table {%
0.8 3
1.2 3
};
\addplot [black, mark=diamond*, mark size=2.5, mark options={solid,fill=darkslategray73}, only marks]
table {%
1 4
1 5
1 4
1 7
1 4
1 5
1 4
1 5
1 5
1 4
1 5
1 5
1 4
1 5
1 4
1 6
1 5
1 4
1 7
1 4
1 4
1 5
1 5
1 5
1 4
1 5
1 4
1 7
1 5
1 5
1 4
1 5
1 5
1 5
1 5
1 5
1 5
1 5
1 5
1 7
1 8
1 5
1 5
1 5
1 7
1 4
1 6
1 5
1 6
1 5
1 4
1 4
1 6
1 5
1 5
1 6
1 5
1 5
1 4
1 4
1 8
1 5
1 5
1 6
1 5
1 6
1 5
1 7
1 5
1 5
1 10
1 6
1 6
1 7
1 5
1 5
1 5
1 6
1 7
1 5
1 6
1 6
1 6
1 6
1 5
1 7
1 5
1 6
1 5
1 5
1 4
1 6
1 5
1 6
1 4
1 9
1 6
1 6
1 4
1 9
1 6
1 6
1 8
1 8
1 9
1 5
1 6
1 9
1 5
1 6
1 17
1 6
1 6
1 10
1 5
1 6
1 10
1 4
1 6
1 6
1 5
1 5
1 9
1 5
1 5
1 9
1 5
1 5
1 9
1 6
1 5
1 6
1 5
1 6
1 9
1 6
1 6
1 9
1 6
1 5
1 9
1 6
1 5
1 9
1 5
1 6
1 5
1 6
1 9
1 5
1 5
1 9
1 6
};
\addplot [semithick, darkslategray73]
table {%
2 1
2 1
};
\addplot [semithick, darkslategray73]
table {%
2 2
2 3
};
\addplot [semithick, darkslategray73]
table {%
1.8 1
2.2 1
};
\addplot [semithick, darkslategray73]
table {%
1.8 3
2.2 3
};
\addplot [black, mark=diamond*, mark size=2.5, mark options={solid,fill=darkslategray73}, only marks]
table {%
2 6
2 4
2 4
2 6
2 4
2 4
2 4
2 4
2 4
2 6
2 4
2 4
2 4
2 5
2 4
2 5
2 5
2 5
2 5
2 7
2 6
2 4
2 4
};
\addplot [semithick, darkslategray73]
table {%
3 1
3 1
};
\addplot [semithick, darkslategray73]
table {%
3 1
3 1
};
\addplot [semithick, darkslategray73]
table {%
2.8 1
3.2 1
};
\addplot [semithick, darkslategray73]
table {%
2.8 1
3.2 1
};
\addplot [black, mark=diamond*, mark size=2.5, mark options={solid,fill=darkslategray73}, only marks]
table {%
3 2
3 3
3 3
3 2
3 2
3 3
3 2
3 3
3 3
3 2
3 2
3 4
3 3
3 6
3 3
3 4
3 4
3 4
3 3
3 2
3 2
3 2
3 3
3 4
3 2
3 3
3 3
3 2
3 4
3 2
3 2
3 3
3 2
3 4
3 3
3 2
3 2
3 3
3 2
3 3
3 3
3 2
3 2
3 2
3 2
3 2
3 2
3 3
3 3
3 3
3 2
3 2
3 2
3 3
3 2
3 2
3 3
3 4
3 2
3 2
3 2
3 2
3 2
3 3
};
\addplot [semithick, darkslategray73]
table {%
4 1
4 1
};
\addplot [semithick, darkslategray73]
table {%
4 1
4 1
};
\addplot [semithick, darkslategray73]
table {%
3.8 1
4.2 1
};
\addplot [semithick, darkslategray73]
table {%
3.8 1
4.2 1
};
\addplot [black, mark=diamond*, mark size=2.5, mark options={solid,fill=darkslategray73}, only marks]
table {%
4 2
4 2
4 3
4 3
4 6
4 2
4 3
4 2
4 3
4 2
4 3
4 7
4 3
4 3
4 2
4 2
4 3
4 6
4 2
4 4
4 2
4 2
4 2
4 2
4 3
4 2
4 2
4 3
4 3
4 3
4 2
4 2
4 2
4 3
4 2
4 4
4 2
4 2
4 2
4 2
4 3
4 2
4 3
4 4
4 6
4 3
4 2
4 3
4 2
4 3
4 2
4 3
4 2
4 3
4 2
4 2
4 2
4 3
4 4
4 3
4 3
4 3
4 2
4 3
4 2
4 2
4 4
4 2
4 3
4 2
4 3
4 2
4 2
4 2
4 2
4 3
4 2
4 2
4 2
4 2
4 2
4 2
4 4
4 2
4 2
4 3
4 3
4 2
4 2
4 2
4 2
4 2
4 3
4 4
4 2
4 2
4 3
4 2
4 2
4 2
4 2
4 3
4 3
4 2
4 3
4 4
4 2
4 2
4 3
4 2
4 4
4 2
4 3
4 2
4 2
4 3
4 3
4 3
4 2
};
\addplot [semithick, darkslategray73]
table {%
5 1
5 1
};
\addplot [semithick, darkslategray73]
table {%
5 1
5 1
};
\addplot [semithick, darkslategray73]
table {%
4.8 1
5.2 1
};
\addplot [semithick, darkslategray73]
table {%
4.8 1
5.2 1
};
\addplot [black, mark=diamond*, mark size=2.5, mark options={solid,fill=darkslategray73}, only marks]
table {%
5 3
5 3
5 3
5 3
5 7
5 6
5 2
5 5
5 4
5 3
};
\addplot [semithick, darkslategray73]
table {%
6 1
6 1
};
\addplot [semithick, darkslategray73]
table {%
6 1
6 1
};
\addplot [semithick, darkslategray73]
table {%
5.8 1
6.2 1
};
\addplot [semithick, darkslategray73]
table {%
5.8 1
6.2 1
};
\addplot [black, mark=diamond*, mark size=2.5, mark options={solid,fill=darkslategray73}, only marks]
table {%
6 8
6 2
6 3
6 3
6 3
6 2
6 3
6 2
6 2
};
\addplot [semithick, darkslategray73]
table {%
7 1
7 1
};
\addplot [semithick, darkslategray73]
table {%
7 5
7 7
};
\addplot [semithick, darkslategray73]
table {%
6.8 1
7.2 1
};
\addplot [semithick, darkslategray73]
table {%
6.8 7
7.2 7
};
\addplot [semithick, darkslategray73]
table {%
-0.4 1
0.4 1
};
\addplot [semithick, darkslategray73]
table {%
0.6 1
1.4 1
};
\addplot [semithick, darkslategray73]
table {%
1.6 1
2.4 1
};
\addplot [semithick, darkslategray73]
table {%
2.6 1
3.4 1
};
\addplot [semithick, darkslategray73]
table {%
3.6 1
4.4 1
};
\addplot [semithick, darkslategray73]
table {%
4.6 1
5.4 1
};
\addplot [semithick, darkslategray73]
table {%
5.6 1
6.4 1
};
\addplot [semithick, darkslategray73]
table {%
6.6 2
7.4 2
};
\end{axis}

\end{tikzpicture}
\caption{Average Packet Distribution per device (Table~\ref{tab0}) in the unit of time, considering only active 1-second windows.}
\label{fig:pck_dev}
\end{figure}

\subsection{Performance Evaluation}
\label{section:perf_eval}
Tests were conducted on real hardware devices that can be used in the field to validate the tool's performance. The test setup involved installing the IoTS tool as an integration on the Home Assistant OS, which was installed on a Raspberry PI 3 Model B with a 1.2 GHz CPU and 1 GB of RAM. A dongle based on the CC2531, a USB-enabled SoC solution for 2.4 GHz IEEE 802.15.4, was used as a radio interface. The dongle was installed with a firmware version developed by Texas Instruments, which enables the Radio Frequency Monitor mode of the CC2531 over serial communication. A Zigbee network was established using an EFR32MG21-based USB dongle as the PANC and several commercially available CIoT devices listed in Table~\ref{tab0}.
Different experiments were performed by rerunning traffic capture and feature extraction to evaluate the tool's performance under different stress conditions. Figure~\ref{fig:cpufull} illustrates the CPU consumption as a function of the packet rate when performing: (i) 5-second time windowed feature extraction with all features selected, (ii) Standard PCAP capture, (iii) Traffic Capture without further processing the packets, to test the acquisition library performance. The system performance was acquired using a bash script, which stores the output of the Linux TOP command as a CSV file. A Crossbow TelosB (TPR2400) mote was used with custom firmware designed to generate 20-byte length IEEE 802.15.4 packets with different destination addresses at a controlled rate to generate additional traffic for IoTS to process, allowing for IoTS tool's performance evaluation under stress condition. To establish a correlation between the performance of the IoTS tool and a real-life scenario, an initial examination of CIoT's traffic consumption was conducted. Figure~\ref{fig:pck_dev} illustrates the packet distribution per device when in an active state. Specifically, the analysis focuses on devices that actively transmit packets over the network at each time interval.
The plot shows that each device has a median of 1 packet per second and many expected outliers, representing the packets exchanged during human-device interaction. The packet rate produced in an LR-WPAN is a function of the number of devices and the number of human-device interactions in the unit of time, which can be considered a sparse event. Experimentally, averaging the number of packets acquired over the entire acquisition time, a packet transmission rate of 4 packets per second was observed. Figure~\ref{fig:cpufull} shows a linear dependency between CPU consumption and Packet rate. With a Packet rate of 40 packets per second (one order of magnitude bigger than the average packet rate generated by eight devices), the CPU consumption presents only an increase of 2.3 \% for the Time Windowed Feature Extraction process. No significant changes were noticed concerning RAM consumption. When comparing the performance metrics of our tool against those of FeatureSniffer\cite{FABIO_OWRT},
our tool's performance exhibits a performance deficit of fivefold. However, some considerations are needed:
Firstly, the two tools are designed to analyze different types of network traffic, leading to different features and packet rates. Secondly, IoTS uses different network hardware interfaces that are unoptimized on both the library and hardware side for network monitoring. Thirdly, IoTS is written in Python, which is slower than unoptimized C by a factor of 45; the choice of programming language was motivated by the integration requirements with the HomeAssistantOS platform, which is fully written in Python.
\section{Use Case} \label{c:use_case}
To demonstrate the insights introduced by IoTS in IoT forensics analyses, this chapter showcases the ability to perform device identification based on IoT traffic features extracted by IoTS. Device identification refers to the ability to recognize a device (brand and device type) by analyzing only some features produced by network traffic. Device identification is a primary and fundamental task in many forensic analyses, as it allows investigators to differentiate between traces left by different devices, which is essential to attribute a particular action or activity to a specific device or user since CIoT devices have a limited range of possible actions. Although previous works have already showcased the Device Identification tasks on Zigbee Network \cite{Acar2020},\cite{Babun2020}, the aims of this work are: (i) Showcase the IoTS tool to extract features from a Zigbee Network and use them directly for the Device Identification task, (ii) Identifying and minimizing the number of features essential for the Device Identification task, (iii) Compare multiple Machine Learning (ML) classification algorithms' performance.
\subsection{Methods}
\begin{table}
\centering
\caption{Features importance ranking from forest of trees classification algorithm}
\label{tab1}
\begin{tabular}{|c|c|} \hline \textbf{Features}&\textbf{Importance Score} \\ \hline
Mean Inter-arrival Time & 0.14\\ \hline
Mean Outgoing Packet Length& 0.10\\ \hline
Mean Incoming Packet Length& 0.09\\ \hline
Mean Incoming Inter-arrival Time & 0.08\\ \hline
Mean Packet Length& 0.07\\ \hline
Mean Outgoing Packet Payload Length & 0.06\\ \hline
\end{tabular}
\end{table}

The data collection phase involved the Zigbee network previously described in Section \ref{section:perf_eval} and the devices in Table~\ref{tab0}. The acquisition process was performed by IoTS using a 5-second time window, selecting all the features described in Section \ref{c:tool}. The traffic was recorded for the duration of five hours under conditions that simulate the typical daily usage of CIoT devices. This was accomplished by simulating the user's interaction with the various IoT devices, including turning the smart bulb on and off, changing its brightness, activating the motion and door sensors, turning the smart outlets on and off, interacting with the wall switch, where for each button pressure was assigned a set of device commands (i.e. toggle on and off the light and the smart outlet) as it would be in a smart home environment.
Before starting to analyze the produced dataset\footnote[4]{https://github.com/antonio-boiano/IoTScent/tree/main/Dataset}, data cleaning and feature selection were performed. This step is fundamental since it improves model performance, provides better model explainability, and reduces the dataset dimensionality, translating into lighter and less complex models. As a result, a dataset containing 24,000 entries was obtained after data cleaning. The Feature Importance was computed for the entire acquired dataset using a Random Forest (RF) Classifier. Forest of trees algorithm works by constructing an ensemble of decision trees, where each tree is trained on a different subset of data. Therefore, it is possible to compute the feature importance based on how much each feature contributes to the overall performance of the ensemble. In our analysis, only features with an importance score greater than 0.06 were deemed suitable and selected for the analysis step. The complete list of features used for the classification algorithms analysis is reported in Table \ref{tab1}.

The performance of popular ML models widely used in literature to solve device identification was evaluated. Among all the models used for the device identification task, we considered mainly decision tree-based ML algorithms since the dataset presents an unbalanced data distribution, with some devices performing periodic transmissions while others producing traffic only when triggered. We evaluated: Decision Tree, XGBoost, Random Forest classifiers, and K-nearest neighbours. The ML algorithms were trained on the features extracted by IoTS described in Table \ref{tab1}. To obtain more reliable and unbiased estimates of model performance, 10-fold cross-validation was employed, which involves dividing the data into 10 non-overlapping folds, using 1 fold for testing and the remaining 9 for training. For evaluating the classification model's performance, the F1 score was used to address unbalanced data distribution.
To synthesize the result among the different classes, the F1 Score macro average is computed, where the Per-Device F1 Score is averaged by performing the arithmetic mean. By doing so, all classes are treated equally regardless of their support values. Moreover, to improve model performance, multiple time window samples can be grouped together based on the device address and the probability of the estimations, averaged for each class label.

\subsection{Results}
\begin{figure}
\centering
\begin{tikzpicture}[scale=0.80]

\definecolor{darkslategray38}{RGB}{38,38,38}
\definecolor{lightgray204}{RGB}{204,204,204}
\definecolor{orange}{RGB}{255,165,0}
\definecolor{yellow}{RGB}{255,255,0}

\begin{axis}[
axis line style={lightgray204},
legend cell align={left},
legend style={
  fill opacity=0.8,
  draw opacity=1,
  text opacity=1,
  at={(0.03,0.97)},
  anchor=north west,
  draw=lightgray204,
  font=\footnotesize
},
ytick pos=left,
xtick pos=bottom,
tick align=outside,
x grid style={lightgray204},
xlabel=\textcolor{darkslategray38}{\(\displaystyle{Number\ of\ Observations\ used}\)},
xmajorticks=true,
xmin=0.0499999999999999, xmax=20.95,
xtick style={color=darkslategray38},
y grid style={lightgray204},
ylabel=\textcolor{darkslategray38}{\(\displaystyle {F1-Score\ Macro\ Averaged}\)},
ymajorgrids,
ymajorticks=true,
ymin=0.768160076731183, ymax=0.907592851879461,
ytick style={color=darkslategray38}
]
\path [draw=orange, fill=orange, opacity=0.2]
(axis cs:1,0.820176637439253)
--(axis cs:1,0.82438955583429)
--(axis cs:2,0.827368424355365)
--(axis cs:3,0.831072425101134)
--(axis cs:4,0.83619622622124)
--(axis cs:5,0.841046319723978)
--(axis cs:6,0.845734169948029)
--(axis cs:7,0.850487259891043)
--(axis cs:8,0.854923029742853)
--(axis cs:9,0.858298502614195)
--(axis cs:10,0.862830846211535)
--(axis cs:11,0.86588880021082)
--(axis cs:12,0.868541337377719)
--(axis cs:13,0.870974657373017)
--(axis cs:14,0.876007390018196)
--(axis cs:15,0.877018339786453)
--(axis cs:16,0.879921974962803)
--(axis cs:17,0.882530177883144)
--(axis cs:18,0.884258042886135)
--(axis cs:19,0.886346975930133)
--(axis cs:20,0.887826188569381)
--(axis cs:20,0.886159352045329)
--(axis cs:20,0.886159352045329)
--(axis cs:19,0.884346945852604)
--(axis cs:18,0.882186502863219)
--(axis cs:17,0.879186120713453)
--(axis cs:16,0.87709807331212)
--(axis cs:15,0.875371354696991)
--(axis cs:14,0.872492091890846)
--(axis cs:13,0.869066387464207)
--(axis cs:12,0.86596820920284)
--(axis cs:11,0.862714532597892)
--(axis cs:10,0.859107988745902)
--(axis cs:9,0.855651550401749)
--(axis cs:8,0.849822629166602)
--(axis cs:7,0.846077052281033)
--(axis cs:6,0.841908978244412)
--(axis cs:5,0.836860051653088)
--(axis cs:4,0.8322999031898)
--(axis cs:3,0.828855349556121)
--(axis cs:2,0.823425638509229)
--(axis cs:1,0.820176637439253)
--cycle;

\path [draw=yellow, fill=yellow, opacity=0.2]
(axis cs:1,0.774497930147014)
--(axis cs:1,0.7800176485824)
--(axis cs:2,0.784782004446294)
--(axis cs:3,0.792042992158293)
--(axis cs:4,0.797319620439273)
--(axis cs:5,0.805141599677055)
--(axis cs:6,0.809804701183322)
--(axis cs:7,0.815099094797906)
--(axis cs:8,0.81769247692201)
--(axis cs:9,0.824598900567406)
--(axis cs:10,0.828471671196277)
--(axis cs:11,0.830094746509618)
--(axis cs:12,0.833732887875017)
--(axis cs:13,0.838010785920094)
--(axis cs:14,0.840822793158178)
--(axis cs:15,0.843506890045793)
--(axis cs:16,0.844551535866616)
--(axis cs:17,0.84807143270848)
--(axis cs:18,0.849763236271824)
--(axis cs:19,0.851657083331812)
--(axis cs:20,0.852505969743462)
--(axis cs:20,0.845636170712618)
--(axis cs:20,0.845636170712618)
--(axis cs:19,0.84191059990474)
--(axis cs:18,0.841879269272819)
--(axis cs:17,0.840966496391895)
--(axis cs:16,0.837276741318677)
--(axis cs:15,0.835050359242621)
--(axis cs:14,0.833478360471668)
--(axis cs:13,0.828325268345208)
--(axis cs:12,0.82723594563371)
--(axis cs:11,0.825293112463078)
--(axis cs:10,0.82206459479625)
--(axis cs:9,0.816181733127117)
--(axis cs:8,0.814980883856821)
--(axis cs:7,0.80934654418184)
--(axis cs:6,0.803448920563313)
--(axis cs:5,0.799929554181463)
--(axis cs:4,0.793759800530297)
--(axis cs:3,0.786361606326928)
--(axis cs:2,0.780976749884357)
--(axis cs:1,0.774497930147014)
--cycle;

\path [draw=red, fill=red, opacity=0.2]
(axis cs:1,0.832065170868178)
--(axis cs:1,0.833986735291878)
--(axis cs:2,0.838468922161955)
--(axis cs:3,0.844053319944255)
--(axis cs:4,0.849645177120751)
--(axis cs:5,0.855278761526951)
--(axis cs:6,0.861533393547821)
--(axis cs:7,0.864204628166654)
--(axis cs:8,0.869000319351083)
--(axis cs:9,0.873274295790153)
--(axis cs:10,0.876055196344085)
--(axis cs:11,0.879846505452392)
--(axis cs:12,0.88336331740194)
--(axis cs:13,0.885500181846541)
--(axis cs:14,0.888781146545667)
--(axis cs:15,0.890556005237838)
--(axis cs:16,0.892779384552352)
--(axis cs:17,0.89644260431508)
--(axis cs:18,0.896533055608934)
--(axis cs:19,0.900445299904024)
--(axis cs:20,0.90125499846363)
--(axis cs:20,0.89785540757168)
--(axis cs:20,0.89785540757168)
--(axis cs:19,0.896884687475188)
--(axis cs:18,0.895557341541848)
--(axis cs:17,0.892393565260394)
--(axis cs:16,0.888983919517267)
--(axis cs:15,0.887804840305745)
--(axis cs:14,0.885880548167192)
--(axis cs:13,0.881544343199854)
--(axis cs:12,0.878680640711649)
--(axis cs:11,0.876310286217539)
--(axis cs:10,0.87242544821427)
--(axis cs:9,0.869105997775169)
--(axis cs:8,0.865501521574923)
--(axis cs:7,0.861360697830487)
--(axis cs:6,0.856136254421589)
--(axis cs:5,0.851651852816889)
--(axis cs:4,0.846850935335912)
--(axis cs:3,0.841448688051312)
--(axis cs:2,0.836916315892134)
--(axis cs:1,0.832065170868178)
--cycle;

\path [draw=blue, fill=blue, opacity=0.2]
(axis cs:1,0.829667525010965)
--(axis cs:1,0.832690141651101)
--(axis cs:2,0.838410129450151)
--(axis cs:3,0.843701548801915)
--(axis cs:4,0.849041002059886)
--(axis cs:5,0.854960326611931)
--(axis cs:6,0.859649578752486)
--(axis cs:7,0.863503118301454)
--(axis cs:8,0.867901479676467)
--(axis cs:9,0.872830984374572)
--(axis cs:10,0.875967135146452)
--(axis cs:11,0.879269419174802)
--(axis cs:12,0.882852853330947)
--(axis cs:13,0.883646497527498)
--(axis cs:14,0.888774710479654)
--(axis cs:15,0.890716865198128)
--(axis cs:16,0.89223601758738)
--(axis cs:17,0.896944986258358)
--(axis cs:18,0.896215232343844)
--(axis cs:19,0.900054508325604)
--(axis cs:20,0.899495284453357)
--(axis cs:20,0.897296166895166)
--(axis cs:20,0.897296166895166)
--(axis cs:19,0.894412384004299)
--(axis cs:18,0.891935288276476)
--(axis cs:17,0.890254921243996)
--(axis cs:16,0.887971281028711)
--(axis cs:15,0.885008490440229)
--(axis cs:14,0.883359844320953)
--(axis cs:13,0.880326519798771)
--(axis cs:12,0.876780402673746)
--(axis cs:11,0.873968526510313)
--(axis cs:10,0.87082330688486)
--(axis cs:9,0.867030164963777)
--(axis cs:8,0.864420811955013)
--(axis cs:7,0.859831422850612)
--(axis cs:6,0.853696451647068)
--(axis cs:5,0.849492087567335)
--(axis cs:4,0.844211512653085)
--(axis cs:3,0.838386029920726)
--(axis cs:2,0.83437074691131)
--(axis cs:1,0.829667525010965)
--cycle;

\addplot [semithick, orange]
table {%
1 0.822283096636771
2 0.825397031432297
3 0.829963887328627
4 0.83424806470552
5 0.838953185688533
6 0.843821574096221
7 0.848282156086038
8 0.852372829454728
9 0.856975026507972
10 0.860969417478718
11 0.864301666404356
12 0.86725477329028
13 0.870020522418612
14 0.874249740954521
15 0.876194847241722
16 0.878510024137462
17 0.880858149298298
18 0.883222272874677
19 0.885346960891369
20 0.886992770307355
};
\addlegendentry{Decision Tree}
\addplot [semithick, yellow]
table {%
1 0.777257789364707
2 0.782879377165325
3 0.78920229924261
4 0.795539710484785
5 0.802535576929259
6 0.806626810873317
7 0.812222819489873
8 0.816336680389415
9 0.820390316847261
10 0.825268132996263
11 0.827693929486348
12 0.830484416754363
13 0.833168027132651
14 0.837150576814923
15 0.839278624644207
16 0.840914138592646
17 0.844518964550187
18 0.845821252772322
19 0.846783841618276
20 0.84907107022804
};
\addlegendentry{KNN}
\addplot [semithick, red]
table {%
1 0.833025953080028
2 0.837692619027045
3 0.842751003997783
4 0.848248056228332
5 0.85346530717192
6 0.858834823984705
7 0.862782662998571
8 0.867250920463003
9 0.871190146782661
10 0.874240322279177
11 0.878078395834965
12 0.881021979056795
13 0.883522262523197
14 0.88733084735643
15 0.889180422771791
16 0.890881652034809
17 0.894418084787737
18 0.896045198575391
19 0.898664993689606
20 0.899555203017655
};
\addlegendentry{Random Forest}
\addplot [semithick, blue]
table {%
1 0.831178833331033
2 0.83639043818073
3 0.841043789361321
4 0.846626257356485
5 0.852226207089633
6 0.856673015199777
7 0.861667270576033
8 0.86616114581574
9 0.869930574669174
10 0.873395221015656
11 0.876618972842557
12 0.879816628002346
13 0.881986508663135
14 0.886067277400304
15 0.887862677819179
16 0.890103649308045
17 0.893599953751177
18 0.89407526031016
19 0.897233446164952
20 0.898395725674262
};
\addlegendentry{XGBoost}
\end{axis}

\end{tikzpicture}
\caption{F1-score evaluation with varying number of observations} 
\label{fig:zb}
\end{figure}
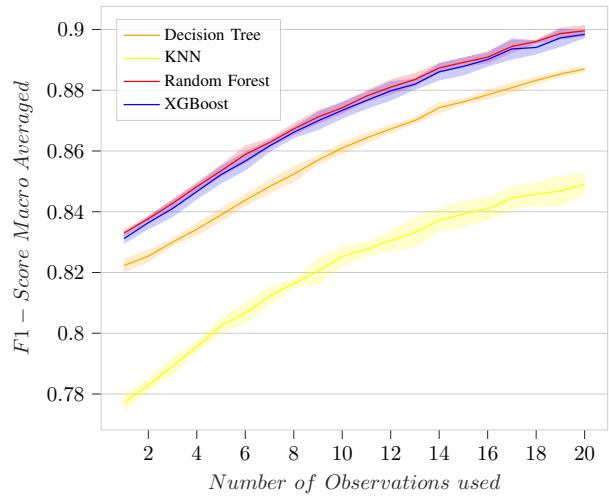
The feature importance described in Table \ref{tab1} identifies the \textit{mean IAT} feature as the one with the highest normalized feature importance score (0.14). This is expected since, as described by \cite{Babun2020}, the features extracted from IAT alone can successfully be used for accurate device identification. This is mainly due to the fact that each device presents a unique IAT fingerprint. Figure~\ref{fig:zb} depicts the macro averaged F1 Score of different ML Classification models in relation to the number of time windows considered. Combining the observations shows an increase in model performance for all the classification algorithms under analysis. Among the analysed models, the ones reporting the highest F1 score are RF, with an F1 score of 0.899 and XGBoost, with an F1 score of 0.898. The RF model performance per device are shown in Table \ref{tab2}
\begin{table}[h]
\centering
\caption{Precision, Recall, and F1-Score for each Device from the RF model with 20 observations combined}
\begin{tabular}{lccc}
\hline
\textbf{Device} & \textbf{Precision} & \textbf{Recall} & \textbf{F1-Score} \\ \hline
PANC & 0.96 & 0.91 & 0.93 \\
(a) & 0.99 & 0.87 & 0.93 \\
(b) & 0.92 & 0.87 & 0.90 \\
(c) & 0.97 & 0.75 & 0.85 \\
(d) & 0.90 & 0.76 & 0.83 \\
(e) & 0.96 & 0.94 & 0.95 \\
(f) & 0.96 & 1.00 & 0.98 \\
(g) & 0.66 & 0.87 & 0.75 \\
Unlabelled & 0.99 & 1.00 & 1.00 \\ \hline
\textbf{Accuracy} & & & 0.90 \\
\textbf{Macro Avg.} & 0.92 & 0.89 & 0.90 \\
\textbf{Weighted Avg.} & 0.91 & 0.90 & 0.90 \\ \hline
\end{tabular}
\label{tab2}
\end{table}
The model performance achieved with our testbed coincides with the ones achieved in literature \cite{Babun2020},\cite{Acar2020}. However, different from these works, the analysis focused on identifying a small subset of features that can be computed online with high efficiency without analysing the encrypted traffic. This could potentially yield advantages in ML model size and complexity, data storage requirements, and CPU computational consumption.
Moreover, by combining multiple observations, the model performance has been increased by 8\%. Nevertheless, the proposed analysis presents some limitations and concerns:
\begin{enumerate}
    \item It is worth mentioning that IoTS alone has no PANC capabilities and can perform packet sniffing in monitor mode (capture all wireless packets within its range, regardless of the target device's address and network). This means that anyone in reach of the PAN can perform the same analysis as described in this chapter. A solution to this privacy leak was proposed by \cite{Acar2020}, where spoofed traffic was generated to degrade the ML model's performance. This mechanism can be integrated into IoTS for enhanced user privacy at the IoT gateway.
    \item The device identification task has undergone training and testing using the same PAN (same set of devices and PANC).
    Hence, it is crucial to confirm that the trained models can generalize under different PANs with similar device sets but managed by different PANCs.

\end{enumerate}

\section{Conclusions} \label{c:conc}
This paper proposed IoTScent, an IoT forensic tool for capturing and analysing traffic from IEEE802.15.4-based networks. 
The integration of IoTS into the popular Home Assistant platform was demonstrated, presenting the tool's performance, and providing a real use-case scenario targeting device identification.
Future research directions will target the improvement of IoTS computational performance and architecture generalization of the tool to extend the support to the wider range of IoT communication protocols. Data compression algorithms can also be introduced to reduce the space needed to store network traffic features for long-term periods. Furthermore, federated learning-based solutions could be implemented on IoTS to train machine learning models in a distributed manner. This approach could lead to more accurate classification models for forensic analyses resolving the privacy concerns related to centralized training.

\balance
\section*{Acknowledgments}
This study was carried out within the MICS (Made in Italy Circular and Sustainable) Extended Partnership and received funding from Next-Generation EU (Italian PNRR M4 C2, Invest 1.3 – D.D. 1551.11-10-2022, PE00000004). CUP MICS D43C22003120001. Additional fundings received by PRIN project COMPACT, CUP: D53D23001340006.

\bibliographystyle{IEEEtran}
\bibliography{IEEEabrv.bib, main}

\end{document}